# Lattice dynamics, phonon chirality and spin-phonon coupling in 2D itinerant ferromagnet Fe$_3$GeTe$_2$

Luojun Du,[1,†,*] Jian Tang,[2,3,4,†] Yanchong Zhao,[2,3,4,†] Xiaomei Li,[2,3,4] Rong Yang,[2,3,4] Xuerong Hu,[1] Xueyin Bai,[1] Xiao Wang,[2] Kenji Watanabe,[5] Takashi Taniguchi,[5] Dongxia Shi,[2,3,4] Guoqiang Yu,[2,3] Xuedong Bai,[2,3,4] Tawfique Hasan,[6] Guangyu Zhang,[2,3,4,*] Zhipei Sun[1,7,*]

[1]Department of Electronics and Nanoengineering, Aalto University, Tietotie 3, FI-02150, Finland

[2]Institute of Physics and Beijing National Laboratory for Condensed Matter Physics, Chinese Academy of Sciences, Beijing 100190, China

[3]Songshan-Lake Materials Laboratory, Dongguan, Guangdong Province, 523808, China

[4]School of Physical Sciences, University of Chinese Academy of Science, Beijing 100190, China

[5]National Institute for Materials Science, 1-1 Namiki, Tsukuba 305-0044, Japan

[6]Cambridge Graphene Centre, University of Cambridge, Cambridge, CB3 0FA, UK

[7]QTF Centre of Excellence, Department of Applied Physics, Aalto University, FI-00076 Aalto, Finland

[†]These authors contributed equally to this work

*Corresponding authors: luojun.du@aalto.fi; gyzhang@aphy.iphy.ac.cn; zhipei.sun@aalto.fi

**Fe$_3$GeTe$_2$ has emerged as one of the most fascinating van der Waals crystals due to its two-dimensional (2D) itinerant ferromagnetism, topological nodal lines and Kondo lattice behavior. However, lattice dynamics, chirality of phonons and spin-phonon coupling in this material, which set the foundation for these exotic phenomena, have remained unexplored. Here we report the first experimental investigation of the phonons and mutual interactions between spin and lattice degrees of freedom in few-layer Fe$_3$GeTe$_2$. Our results elucidate three prominent Raman modes at room temperature: two A$_{1g}$(Γ) and one E$_{2g}$(Γ) phonons. The doubly degenerate E$_{2g}$(Γ) mode reverses the helicity of incident photon, indicating the pseudo-angular momentum and chirality. Through analysis of temperature-dependent phonon energies and lifetimes, which strongly diverge from the anharmonic model below Curie temperature, we determine the spin-phonon coupling in Fe$_3$GeTe$_2$. Such interaction between lattice oscillations and spin significantly enhances the Raman susceptibility,**

**allowing us to observe two additional Raman modes at the cryogenic temperature range. In addition, we reveal laser radiation induced degradation of $Fe_3GeTe_2$ in ambient conditions and the corresponding Raman fingerprint. Our results provide the first experimental analysis of phonons in this novel 2D itinerant ferromagnet and their applicability for further fundamental studies and application development.**

## 1. Introduction

Two-dimensional (2D) intrinsic magnetism lies at the heart of both fundamental science and technical applications, such as magneto-electrics, topological magnons, electrical control of spin and ultracompact spintronics.[1-15] In 2017, 2D ferromagnetic order was first demonstrated in $CrI_3$[16] and $Cr_2Ge_2Te_6$,[17] breaking the long-established Mermin-Wagner theorem.[18] Motivated by the discovery of 2D ferromagnetism in these insulating chromium compounds, 2D itinerant ferromagnetic order was recently uncovered in $Fe_3GeTe_2$,[19, 20] offering an unprecedented platform for manipulation of both spin and charge degrees of freedom in the monolayer limit. Moreover, the ferromagnetic transition temperature ($T_c$) of pristine $Fe_3GeTe_2$ is relatively high (~ 200 K),[21-24] and can be tuned to room temperature by electron doping as a result of changes in its electronic structure.[19] In addition to the fascinating 2D itinerant ferromagnetism, $Fe_3GeTe_2$ exhibits a wide range of interesting physical phenomena, such as topological nodal lines,[25] tunneling spin-valve behavior,[26] heavy fermion states,[27] manipulable magnetic domains[28, 29] and strong electron correlation effects.[30] To fully understand these novel phenomena and explore new spin physics, it is critical to investigate the lattice dynamics and mutual coupling between magnetism and energy quanta of lattice oscillations in itinerant ferromagnetic $Fe_3GeTe_2$.[31-43] In addition, because of the hexagonal honeycomb structure and threefold rotational symmetry of $Fe_3GeTe_2$,[25] phonons at the high-symmetry points of Brillouin-zone are expected to possess circular polarization and chirality.[44-47] Therefore, determining the chiral phonons in $Fe_3GeTe_2$ is highly desirable and potentially important for novel phenomena, such as phonon Hall effect[48] and topological states.[49]

In this work, we study the lattice dynamics, chirality of phonons and spin-phonon coupling in $Fe_3GeTe_2$ via systematic Raman measurements. Two $A_{1g}(\Gamma)$ and one $E_{2g}(\Gamma)$ phonons are observed at room temperature. Our results show that the doubly degenerate $E_{2g}(\Gamma)$ mode switches the helicity of incident photons and thus is related to the chiral phonon in $Fe_3GeTe_2$. Moreover, we confirm the spin-phonon coupling effects via a combined analysis of the anomalous phonon energies and linewidths below $T_c$. Such coupling between lattice vibrations and magnetism strongly enhances the Raman susceptibility in $Fe_3GeTe_2$. In addition, two new phonon modes associated with the degradation of $Fe_3GeTe_2$ are uncovered, which could potentially be used as a simple and fast characterization method to determine the quality of $Fe_3GeTe_2$ flakes.

## 2. Results and Discussions

**Selection rules for Raman processes in $Fe_3GeTe_2$.** $Fe_3GeTe_2$ crystallizes in a hexagonal structure and belongs to the $D_{6h}^4$ point group (space group $P6_3$/mmc, No. 194, Figures 1a-1c).[22, 25] According to group theory, the symmetries of Raman active modes and the corresponding Raman tensors ($R$) are described as follows:[50]

$$A_{1g}: \begin{pmatrix} a & 0 & 0 \\ 0 & a & 0 \\ 0 & 0 & b \end{pmatrix}; \quad E_{1g}: \begin{pmatrix} 0 & 0 & 0 \\ 0 & 0 & c \\ 0 & c & 0 \end{pmatrix} \begin{pmatrix} 0 & 0 & -c \\ 0 & 0 & 0 \\ -c & 0 & 0 \end{pmatrix}; \quad E_{2g}: \begin{pmatrix} 0 & d & 0 \\ d & 0 & 0 \\ 0 & 0 & 0 \end{pmatrix} \begin{pmatrix} d & 0 & 0 \\ 0 & -d & 0 \\ 0 & 0 & 0 \end{pmatrix}.$$

Note that $E_{1g}$ and $E_{2g}$ modes are doubly degenerate. For the non-resonant Raman scattering intensity of a Raman-active mode in a crystal, it can be calculated within the Placzek approximation and expressed by the Raman tensor as:[50]

$$I \propto |e_i \cdot R \cdot e_s|^2 \quad (1)$$

where $e_i$ and $e_s$ are polarization vectors of the incident and scattered light, respectively. In the back-scattering configuration, $e_i$ and $e_s$ are within the $xy$ plane. Using the tabulated Raman tensors and Eq. (1), we can determine the selection rules and Raman intensities for various scattering geometries (see Supporting Information for more details). Table 1 summarizes the Raman response in the back-scattering geometry for four polarization configurations. It shows that $A_{1g}(\Gamma)$ and $E_{2g}(\Gamma)$ are Raman active but possess distinct selection rules, while $E_{1g}(\Gamma)$ mode is normally forbidden.

**Lattice dynamics in $Fe_3GeTe_2$.** Having identified the selection rules, we focus on Raman spectra to study the lattice dynamics of $Fe_3GeTe_2$. Few-layer samples were mechanically exfoliated from bulk crystals onto $SiO_2$/silicon substrates (see Experimental Section for more details). Since laser irradiation may lead to the degradation (as discussed in detail below), we covered a thin-layer boron nitride ($h$-BN) on the $Fe_3GeTe_2$ flake before Raman measurements using the dry transfer method in an inert atmosphere. Figure 1d presents the optical microscope image of an exfoliated few-layer $Fe_3GeTe_2$ flake covered by *h*-BN. The corresponding topography from atomic force microscopy (AFM) is shown in Figure 1e. The height profiles (Figure 1f) indicate that the thickness of $Fe_3GeTe_2$ (*h*-BN) is 12.8 (21.3) nm. We performed Raman measurements of $Fe_3GeTe_2$ flakes with difference thicknesses. The results indicate that the 12.8-nm thick $Fe_3GeTe_2$ sample has a relatively strong Raman response and its spectral peak at 120 $cm^{-1}$ can be well distinguished from the peak at 155 $cm^{-1}$ (Figure S2 in Supporting Information). We therefore focus our study on the 12.8 nm thick $Fe_3GeTe_2$ sample.

Figure 1g shows the Raman spectra at room temperature excited by two laser sources at the same condition: 2.33 eV (black) and 1.96 eV (red). Two peaks are clearly resolved at ~ 120 and ~ 155 $cm^{-1}$, respectively. Since phonons (especially for the mode at ~ 120 $cm^{-1}$) have stronger response under 2.33 eV excitation than that at 1.96 eV excitation, all subsequent Raman measurements, unless specified, were performed under 2.33 eV radiation. Figure 1h presents the Raman spectra excited by linearly polarized light and collected in co-polarized (linearly parallel, black) or cross-polarized (linearly perpendicular, red) geometry. The phonon at ~ 120 $cm^{-1}$ can only be observed when the scattered light is parallel to the incident light. We thus deduce that the symmetry of this phonon belongs to $A_{1g}(\Gamma)$ due to the selection rule (Table 1). In contrast, the peak at ~ 155 $cm^{-1}$ can be detected simultaneously in both co-polarized and cross-polarized polarization configurations. Therefore, it should possess a component of $E_{2g}(\Gamma)$. Note that the peak intensity at ~ 155 $cm^{-1}$ under co-polarized configuration is much stronger than that under cross-polarized configuration (Figure 1h). This excludes the possibility that the 155 $cm^{-1}$ peak is composed of only the $E_{2g}(\Gamma)$ vibration, according to the selection rule in Table 1. We speculate that the 155 $cm^{-1}$ peak possesses not only the $E_{2g}(\Gamma)$ mode but also the $A_{1g}(\Gamma)$ phonon. To confirm this hypothesis, we performed helicity-resolved Raman spectra (Figure 1i), which are typically used to assign and resolve the symmetries and chirality of Raman bands.[46, 51-53] Figure 1i shows that the phonon at ~ 120 $cm^{-1}$ appears only in co-circularly polarized geometry ($\sigma^+$ in $\sigma^+$ out), further indicating that the symmetry can be unequivocally assigned to $A_{1g}(\Gamma)$. In marked contrast, the mode at ~ 155 $cm^{-1}$ can be observed in both co- and cross-circularly ($\sigma^+$ in $\sigma^-$ out) polarized configurations, revealing that both $A_{1g}(\Gamma)$ and $E_{2g}(\Gamma)$ are indeed located at ~ 155 $cm^{-1}$.

**Phonon chirality in $Fe_3GeTe_2$.** Strikingly, the helicity of incident photons is completely reversed in the Raman process involving the doubly degenerate $E_{2g}(\Gamma)$ phonons. This phenomenon can be understood as the chirality and pseudo-angular momentum (PAM) of the $E_{2g}(\Gamma)$ modes.[46, 47] Although each single phonon at Brillouin-zone center ($\Gamma$) does not have the intrinsic chirality, we can obtain the chiral phonons via superposing the doubly degenerate $E_{2g}(\Gamma)$ modes.[47] Due to the threefold rotational symmetry, chiral phonons could possess PAM, which consists of two contributions: spin PAM ($l_s$) and orbital PAM ($l_o$). For the doubly degenerate $E_{2g}(\Gamma)$ chiral phonons, the $l_o$ is zero and the total PAM is thus equal to $l_s$ (i.e., 1 or -1). Under the first-order Stokes Raman scattering, the PAM is conserved with optical selection rule of $\Delta l_{photon} = l_s + 3N$, where $\Delta l_{photon}$ is the change of photon angular momentum[47] and $3N$ denotes the azimuthal quantum number of crystal lattice.[54, 55] Therefore, in our experiments, the incident left-circularly polarized photon ($\sigma^+$) emits a right-handed phonon ($l_s = -1$), and then scatters into the right-circularly polarized photon ($\sigma^-$) (inset in Figure 1i).

**Strong spin-phonon coupling in Fe$_3$GeTe$_2$.** After the identification of the lattice dynamics and phonon chirality via polarization-resolved Raman study, we perform temperature-dependent measurements to investigate the spin-phonon coupling, which plays a critical role for many physical phenomena, such as spin-Peierls transition, phonon Hall effect and spin-Seebeck effect.[33-35, 48] Since $A_{1g}(\Gamma)$ at higher energy and $E_{2g}(\Gamma)$ are almost degenerate at ~ 155 cm$^{-1}$ and possess very large linewidths (~ 30 cm$^{-1}$) (Figure S1 and Table S2 in Supporting Information), we focus on the 120 cm$^{-1}$ $A_{1g}(\Gamma)$ mode to study the spin effects. The temperature-driven evolution of phonon frequencies ($\omega$) and linewidths ($\Gamma$) for the $A_{1g}(\Gamma)$ mode at ~ 120 cm$^{-1}$ are presented in Figures 2b and 2d, respectively. Assuming that the system has no spin-phonon coupling, temperature dependent self-energy for the phonon Green's function can be understood as a result of symmetric anharmonic decay, i.e., decay into a pair of acoustic modes with identical frequencies and opposite momenta.[56, 57] The expressions for the frequencies (real part of self-energy) and linewidths (imaginary part of self-energy) can be described in terms of the anharmonic model:

$$\omega_{anh}(T) = \omega_0 - A\left[1 + \frac{2}{(e^{\frac{\hbar\omega_0}{2k_BT}})-1}\right], \qquad (2)$$

$$\Gamma_{anh}(T) = \Gamma_0 + B\left[1 + \frac{2}{(e^{\frac{\hbar\omega_0}{2k_BT}})-1}\right], \qquad (3)$$

where $\omega_0$ is the bare phonon frequency, $\Gamma_0$ denotes the absolute zero-temperature linewidth originating from imperfections or electron-phonon coupling, $k_B$ is the Boltzmann constant, A and B are positive adjustable parameters. As shown in Figure 2b, $\omega$(T) dependence diverges from the standard anharmonic model (red line) and exhibits an anomalous softening at low temperatures. Figure 2c shows the Raman frequency deviation from the anharmonic model $\Delta\omega$(T), defined as $\Delta\omega(T) = \omega(T) - \omega_{anh}(T)$. It shows that the phonon softens quickly below 140 K. Such pronounced softening can be ascribed to spin-phonon coupling caused by the modulation of the super-exchange integral by lattice vibrations.[58-65] From the onset of spin-phonon coupling, the $T_c$ of this 12.8 nm thick $Fe_3GeTe_2$ sample is ~ 140 K, in good agreement with the anomalous Hall effect measurements (Figure S4 in Supporting Information). In a nearest-neighbor approximation, $\Delta\omega$(T) stemmed from the spin-phonon coupling is given as $\Delta\omega(T) = \lambda \langle \mathbf{S}_i \cdot \mathbf{S}_j \rangle$, where $\lambda$ represents the strength of the spin-phonon coupling, and $\langle \mathbf{S}_i \cdot \mathbf{S}_j \rangle$ denotes the spin-spin correlation function of adjacent spins.[35, 65, 66] For $Fe_3GeTe_2$, the saturation moment is about 1.625 $\mu_B$/Fe at 0 K and the corresponding $\langle \mathbf{S}_i \cdot \mathbf{S}_j \rangle$ is ~ 0.66.[22] Considering $\Delta\omega(T) \approx 0.9$ cm$^{-1}$ at 30 K, $\lambda$ can be determined as ~ 1.36 cm$^{-1}$, which is in the same order of magnitude with other 2D magnet, such as $Cr_2Ge_2Te_6$.[35] The mutual interactions between spins and phonons are further confirmed via the lifetime of phonons. As illustrated in Figure 2d, the anharmonic model (red line) fails to describe the $\Gamma$(T) for temperatures below 180 K, which almost becomes a plateau. Figure 2e shows the deviation of linewidths from the anharmonic model $\Delta\Gamma(T) = \Gamma(T) - \Gamma_{anh}(T)$, which increases significantly below 180 K. This indicates that in addition to the anharmonic component, a new relaxation mechanism caused by the spin-phonon coupling has emerged.[37, 67] Strikingly, the temperature, at which the spin-phonon coupling takes effect on the lifetime of phonons, is higher than the $T_c$. This is due to that spin correlations have occurred before the ferromagnetic transition, leading to the magnetic scattering and enhanced linewidths.[37]

**Enhanced optical nonlinearity with spin-phonon coupling.** Further, we observed two new phonon modes around 240 cm$^{-1}$ at cryogenic temperature range (labeled as p1 and p2 in Figure 2a). It is worth noting that the responses of the two modes are weak and their symmetries cannot be determined by polarization-resolved Raman. To elaborate their physical origin, we derive the Raman scattering susceptibility $\chi''(\omega)$ from the raw Raman spectra $I(\omega)$ in Figure 2a using the following relation:[67, 68]

$$\chi''(\omega) = \frac{I(\omega)}{1 + \frac{1}{(e^{\frac{\hbar\omega}{k_B T}})-1}} \qquad (4)$$

Figure 3a presents the temperature dependent $\chi''(\omega)$ (the temperatures of the curves are consistent with that in the same colour in Figure 2a). The $\chi''(\omega)$ increases dramatically at low temperature. Figures 3b and 3c show the temperature-driven evolution of integrated $\chi''(\omega)$ for the 120 cm$^{-1}$ $A_{1g}$ mode (Figure 3b) and 240 cm$^{-1}$ p2 phonon (Figure 3c). $\chi''(\omega)$ is virtually temperature-independent above 140 K and increases dramatically below 140 K. The significantly enhanced $\chi''(\omega)$ below 140 K can be understood as a result of spin-phonon coupling, which gives rise to an increase of Raman polarizability tensor via the variation of electron hopping amplitude with phonons and non-diagonal exchange energy between the magnetic Fe ions.[59, 69, 70] Due to the strongly enhanced $\chi''(\omega)$ below $T_c$, we could observe the phonon modes around 240 cm$^{-1}$.

**Raman fingerprints of degradation in Fe$_3$GeTe$_2$.** Finally, we study the stability of the 2D layered Fe$_3$GeTe$_2$, which could potentially be used to benchmark sample qualities. Figure 4a shows the AFM topography of another freshly exfoliated Fe$_3$GeTe$_2$ flake with thickness of 10.5 nm. The corresponding phase image is presented in Figure 4b. Note that this Fe$_3$GeTe$_2$ flake is not covered by *h*-BN. Interestingly, we found that 2D itinerant ferromagnet Fe$_3$GeTe$_2$ is relatively stable under atmospheric conditions, in marked contrast to other recently discovered 2D ferromagnetic order in CrI$_3$ or Cr$_2$Ge$_2$Te$_6$ which can hardly exist in the atmosphere.[16, 17, 71] This presents an advantage of Fe$_3$GeTe$_2$ over these 2D ferromagnetic orders for further fundamental studies and application development. However, we note that laser radiation can induce degradation in unprotected Fe$_3$GeTe$_2$. Figures 4c and 4d show the AFM height and phase images of Fe$_3$GeTe$_2$ after a series of Raman measurements, indicating the damage for the irradiated part of the sample (dotted red circle). Figure 4e presents a series of Raman spectra measured continuously (see Supporting Information for more details). As the number of Raman measurements increases, phonon modes associated with Fe$_3$GeTe$_2$ (black dotted lines) disappear gradually. At the same time, we could observe two new Raman vibrations, illustrated by the red lines in Figure 4e and located at ~ 122 cm$^{-1}$ and ~ 140 cm$^{-1}$, respectively. Although the assignment of such two new peaks deserves further investigation, the two new phonons associated with the degradation of Fe$_3$GeTe$_2$ provide us a simple and rapid characterization route to determine the quality of the Fe$_3$GeTe$_2$ flake.

## 3. Conclusions

In conclusion, we uncover, for the first time, the lattice dynamics and the chirality of Raman active modes in $Fe_3GeTe_2$ via linear polarization- and helicity-resolved Raman scattering measurements. Two $A_{1g}(\Gamma)$ and one $E_{2g}(\Gamma)$ phonons are observed at room temperature. The doubly degenerate $E_{2g}(\Gamma)$ mode switches the helicity of incident photon and thus possesses chirality. Moreover, our measurements reveal the spin-phonon coupling in $Fe_3GeTe_2$ through the anomalous softening of phonon energies and plateau of linewidths below $T_c$. Such coupling between spin and lattice degrees of freedom gives rise to the significantly enhanced Raman scattering susceptibility, activating new Raman modes at low temperatures. Finally, we identify the Raman fingerprints associated with the degradation of $Fe_3GeTe_2$. Our results provide a firm basis for a cornucopia of peculiar physical phenomena and will prove important in the engineering of spintronic devices.

## 4. Experimental Section

**$Fe_3GeTe_2$ exfoliation.** We adopted a developed scratching scotch tape method to obtain the isolated $Fe_3GeTe_2$ flakes from the bulk materials (HQ Graphene).[72] First, polydimethylsiloxane (PDMS) was used to cleave isolated $Fe_3GeTe_2$ flakes from bulk crystals with the aid of scotch tape. The thickness of the $Fe_3GeTe_2$ layers was identified by optical image and further confirmed by atomic force microscopy. Then, the exfoliated few-layer $Fe_3GeTe_2$ flakes on PDMS were transferred onto a $SiO_2$ substrate through our home-made transfer station in a glove box with $N_2$ atmosphere. By moving the $SiO_2$/Si substrate and PDMS in opposite directions, the isolated $Fe_3GeTe_2$ flakes on PDMS could be successfully transferred onto $SiO_2$/Si substrate. Before transfer, the $SiO_2$/Si substrate was ultrasonically cleaned in acetone, 2-propanol, and deionized (DI) water, and then subjected to oxygen plasma to remove ambient adsorbates on its surface. Next, the $Fe_3GeTe_2$ flake was encapsulated by hBN immediately in order to protect $Fe_3GeTe_2$ oxidization.

**Raman measurements.** Raman spectra were acquired using a micro-Raman spectrometer (Horiba LabRAM HR Evolution) in a confocal backscattering configuration (confocal pinhole of 200 μm). Light from a 532-nm (room temperature) or 633-nm (low temperature) laser was focused down to a 1 μm spot. The laser power on the sample during Raman measurement was kept below 150 μW in order to avoid sample damage and excessive heating. The integration time is 90 s.

For linear polarization- and helicity-resolved Raman spectra at room temperature, the backscattered signal was collected by an Olympus 100× objective (N.A. = 0.95) and dispersed by a 1800-g/mm grating to achieve Raman spectral resolution better than 1 cm$^{-1}$. An initial polarizer sets the linear polarization of the incident light, and a polarizer before the detector controls the polarization of the measured light. For the linear polarization configurations, a half-wave plate in the path of the outgoing light controls which linear polarization is detected. For the helicity-resolved Raman measurement, the excitation laser was first guided through a vertical linear polarizer followed by a quarter-wave plate to achieve σ+ circular polarization. The circular polarization of the excitation light was confirmed at the sample position. The backscattered Raman signal going through the same quarter-wave plate was collected and analyzed with a half-wave plate and a linear polarizer.

**AFM characterizations.** The AFM images were performed by tapping mode AFM (Asylum Research Cypher S) with AC160TS tip at room temperature under ambient condition.

**Scanning transmission electron microscopy (STEM) characterization.** The $Fe_3GeTe_2$ thin flake was first encapsulated in few layer graphene on $SiO_2$ substrate and transferred onto Cu foil by wet method assisted by PMMA and HF solution. The atomic-resolution STEM characterizations were acquired from an aberration-corrected JEOL ARM300F transmission electron microscope which was operated at 80 kV with convergence angle at 18 mrad and collection angles at 54~220 mrad.

**Acknowledgements**

The authors thanks the financial supports from Business Finland (A-Photonics), Academy of Finland (grants: 276376, 284548, 286920, 295777, 298297,304666, 312297, 312551, 314810), Academy of Finland Flagship Programme (320167, PREIN), the European Union's Horizon 2020 research and innovation programme (820423,S2QUIP), NSFC (grants:11834017 and 61888102), the Strategic Priority Research Program of CAS (grant: XDB30000000), the Key Research Program of Frontier Sciences of CAS (grant:QYZDB-SSW-SLH004), the National Key R&D program (grant: 2016YFA0300904). Growth of hexagonal boron nitride crystals was supported by the Elemental Strategy Initiative conducted by the MEXT, Japan and the CREST (JPMJCR15F3),

JST.

**Author contributions**

L.D. and Z.S. designed and supervised the research; L.D. performed the Raman measurements and data analysis; J.T. prepared the samples and performed the AFM characterizations; X.L. and X.B. performed the STEM characterizations; X.W and G.Y performed the anomalous Hall effect measurements; K.W. and T.T. provided the *h*-BN samples; Y.Z., R.Y., X.H., X.B., D.S., G.Y. and G. Z. helped analyzed data; L.D., T.H., G.Z. and Z.S. wrote, and all authors commented on the manuscript.

**Table 1. Selection Rules for Raman-active Phonons**

|  | $A_{1g}(\Gamma)$ | $E_{1g}(\Gamma)$ | $E_{2g}(\Gamma)$ |
|---|---|---|---|
| $I_{\text{linear} \parallel}$ | $|a|^2$ | 0 | $|d|^2$ |
| $I_{\text{linear} \perp}$ | 0 | 0 | $|d|^2$ |
| $I_{\sigma+ \text{ in } \sigma+ \text{ out}}$ | $|a|^2$ | 0 | 0 |
| $I_{\sigma+ \text{ in } \sigma- \text{ out}}$ | 0 | 0 | $2|d|^2$ |

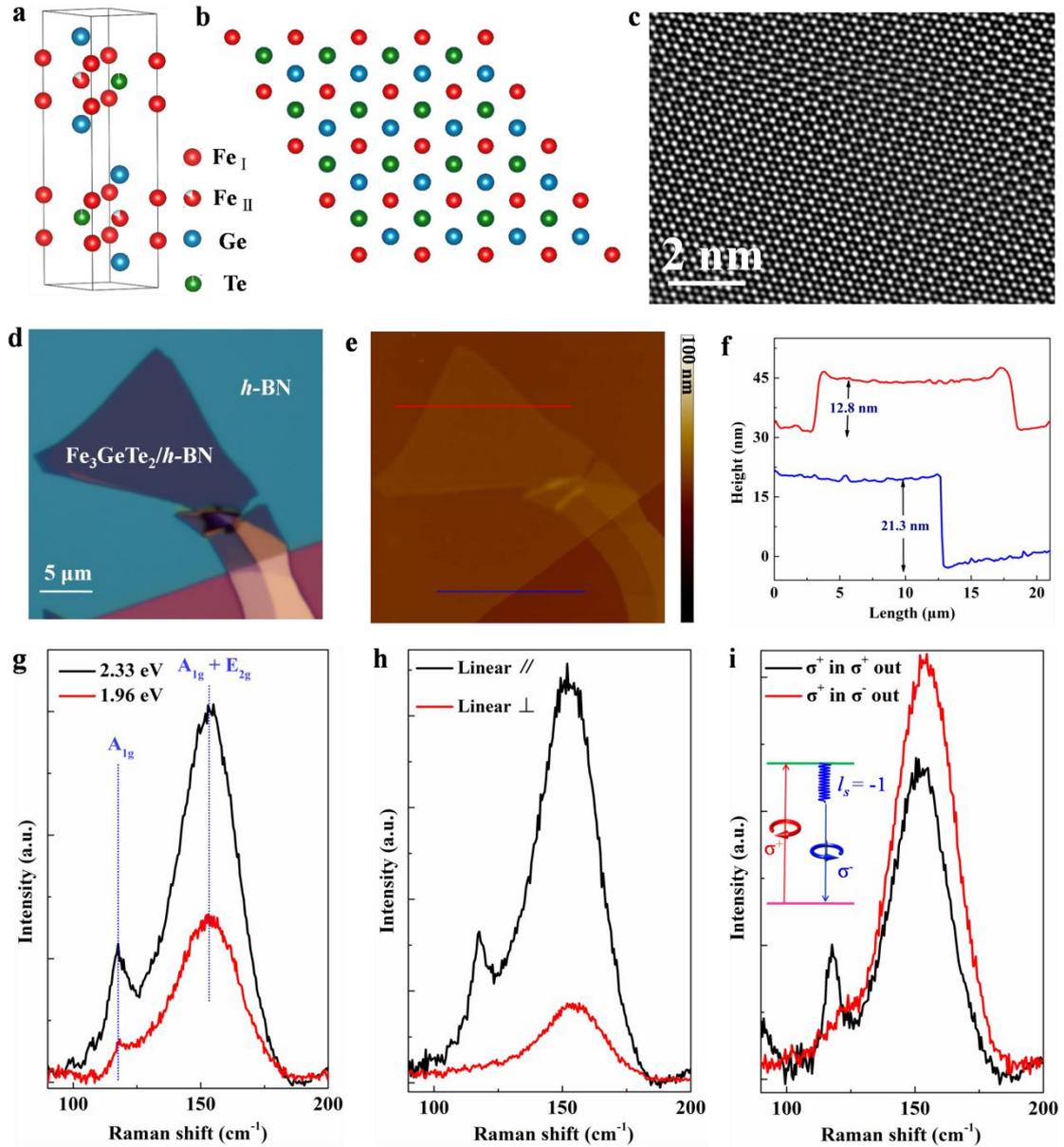

**Figure 1.** Lattice dynamics and the chirality of Raman active phonons of $Fe_3GeTe_2$. (a) Side and (b) top views of the crystal structure of $Fe_3GeTe_2$. (c) Atomic-resolution scanning transmission electron microscopy image. (d) Optical micrograph of a few-layer $Fe_3GeTe_2$ covered with a thin layer of hBN. (e) Atomic force micrograph corresponding to (d). (f) Height profiles taken along the red and blue lines in (e). (g) Raman spectra for two excitation energies: 2.33 eV and 1.96 eV. (h) Linear polarization-resolved Raman spectra. (i) Helicity-resolved Raman spectra. Inset is the schematic of the helicity of photon switched by the chiral $E_{2g}(\Gamma)$ phonon.

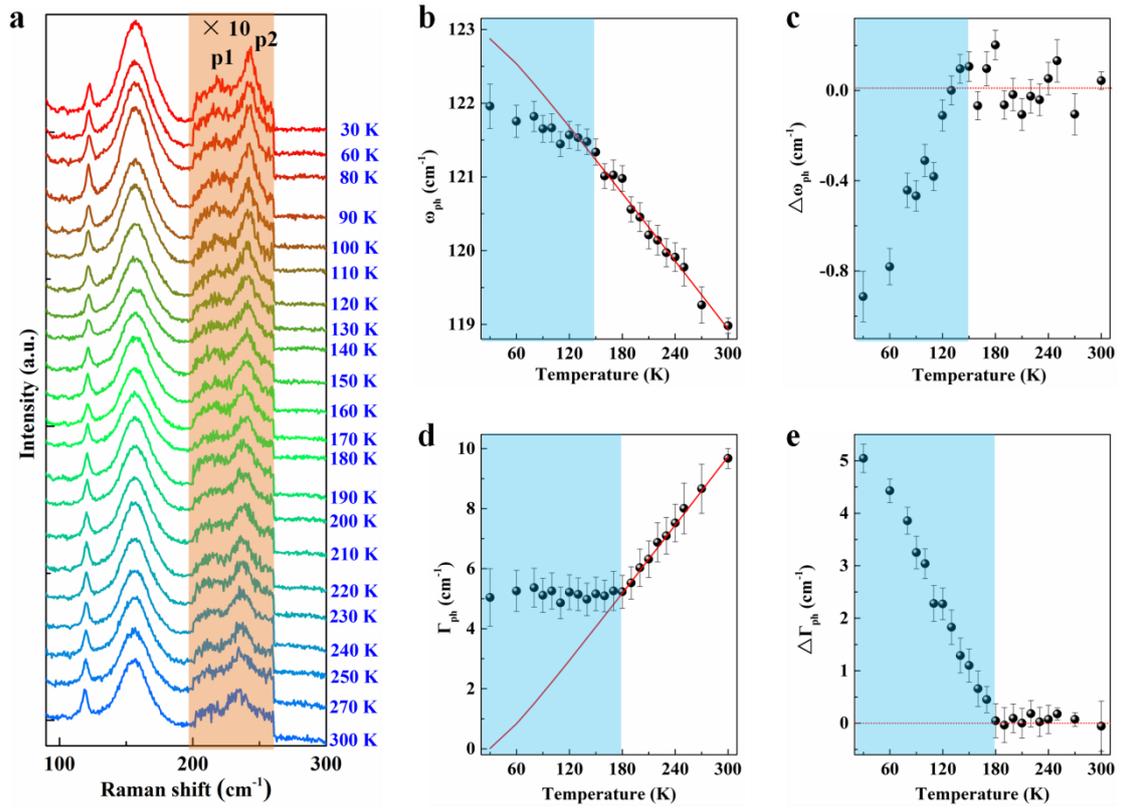

**Figure 2.** Spin-phonon coupling in $Fe_3GeTe_2$. (a) Raman spectra of $Fe_3GeTe_2$ versus temperature. The region highlighted by the red square is magnified 10 times. (b) Phonon energies as a function of temperature. (c) Magnetic-order-induced renormalization of the phonon frequencies. (d) Phonon linewidths versus temperature. (e) Renormalization of the phonon linewidths induced by magnetic ordering. The red curves in (b) and (d) correspond to the results of anharmonic model. The shaded blue region with 70% transparency in (b)- (e) indicates spin-phonon coupling.

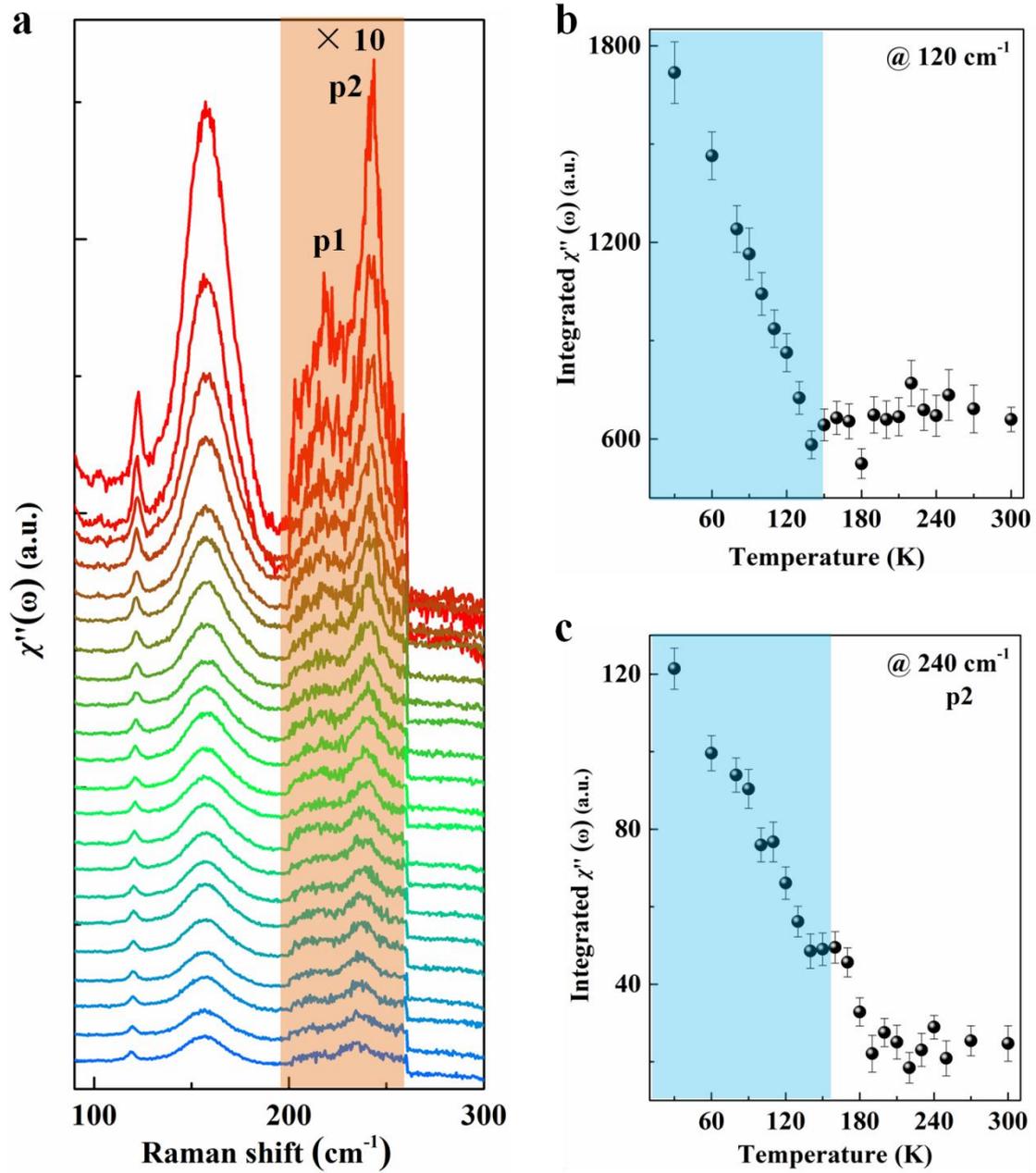

**Figure 3.** Strongly enhanced Raman susceptibility in Fe$_3$GeTe$_2$. (a) Raman scattering susceptibility as a function of temperature. The region highlighted by the red square is magnified 10 times. (b, c) Integrated Raman susceptibility versus temperature for the 120 cm$^{-1}$ A$_{1g}$ mode (b) and 240 cm$^{-1}$ p2 phonon (c). The region highlighted by the blue square with 70% transparency indicates spin-phonon coupling.

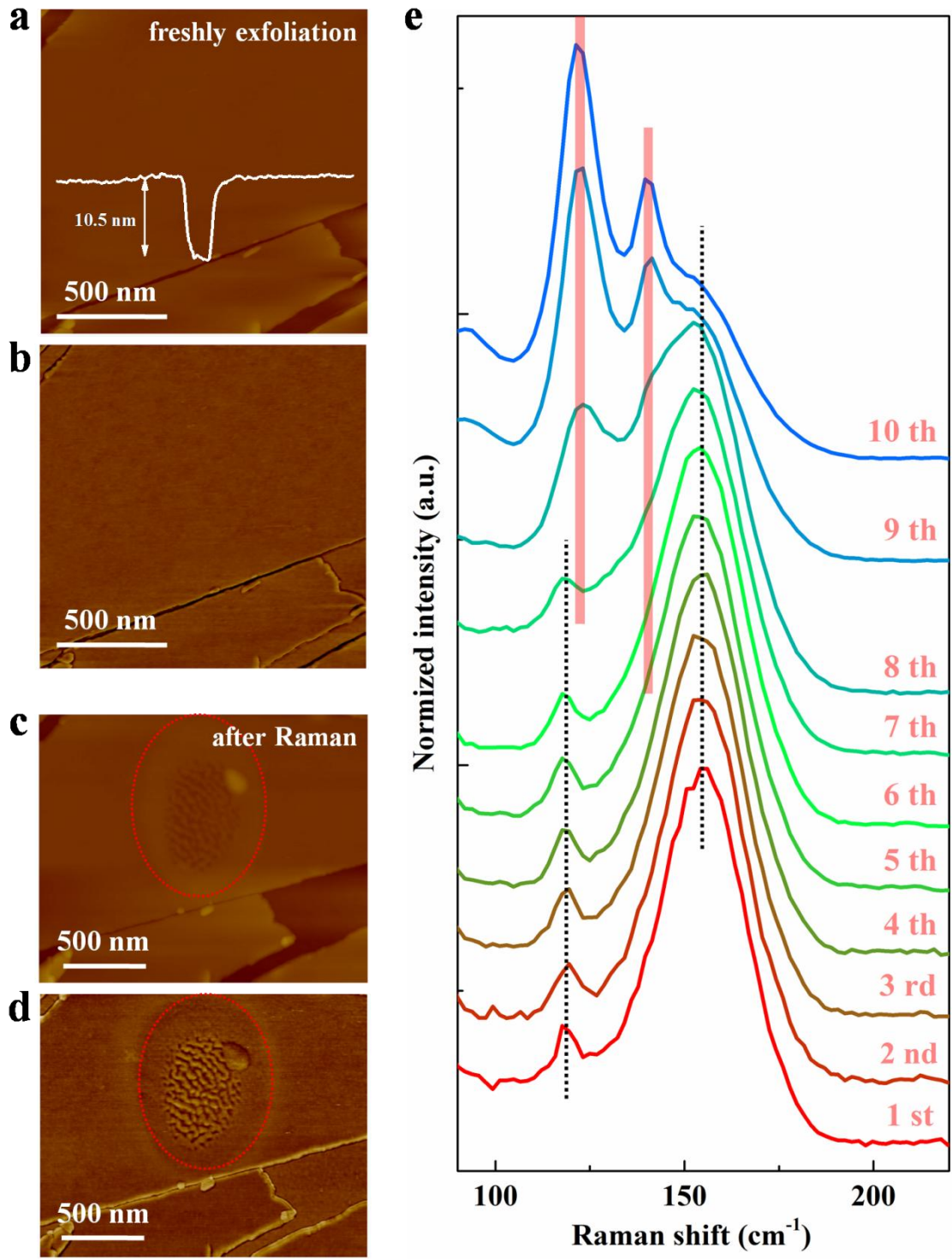

**Figure 4.** Laser radiation-induced degradation in $Fe_3GeTe_2$. (a, b) AFM height images of a freshly exfoliated $Fe_3GeTe_2$ flake. (c, d) AFM height (c) and phase (d) images of $Fe_3GeTe_2$ after a series of Raman measurements. Dotted red circles indicate the position of laser spot. (e) A series of Raman measurements. The black dotted lines indicate the intrinsic phonons of $Fe_3GeTe_2$. The vertical red lines indicate two new Raman modes due to flake degradation.